\documentstyle[aps,pra,epsfig,twocolumn]{revtex}

\def\be{\begin{equation}}
\def\ee{\end{equation}}
\def\bea{\begin{eqnarray}}
\def\eea{\end{eqnarray}}
\def\bma{\begin{mathletters}}
\def\ema{\end{mathletters}}
\def\C{\hbox{$\mit I$\kern-.7em$\mit C$}}
\newcommand{\one}{\mbox{$1 \hspace{-1.0mm}  {\bf l}$}}

\begin{document}
\draft

\title{Catalysis in non--local quantum operations}

\author{G. Vidal and J. I. Cirac}

\address{Institut f\"ur Theoretische Physik, Universit\"at Innsbruck,
A-6020 Innsbruck, Austria}

\date{\today}

\maketitle

\begin{abstract}
We show how entanglement can be used, without being consumed, to
accomplish unitary operations that could not be performed with
out it. When applied to infinitesimal transformations our method
makes equivalent, in the sense of Hamiltonian simulation, a
whole class of otherwise inequivalent two-qubit interactions.
The new catalysis effect also implies the asymptotic equivalence
of all such interactions.

\end{abstract}

\pacs{03.67.-a, 03.65.Bz, 03.65.Ca, 03.67.Hk}

\narrowtext

Can entanglement help to perform certain tasks? How much
entanglement has to be consumed? Can we use entanglement without
consuming it at all? These questions are quite relevant in the
context of quantum information theory, since entanglement can be
considered as an expensive physical resource without classical
analogy. In particular, the last question has been recently
answered \cite{Jo99} in the context of transformation between
states of two parties, Alice and Bob, under local operations and
classical communication (LOCC). More specifically, examples have
been presented where a state can only be transformed into some
other one by LOCC when a certain entangled state
$|\eta\rangle_{ab}$ is available. In this case, even though the
total entanglement (shared by Alice and Bob) decreases, the
state $|\eta\rangle_{ab}$ is recovered after the procedure. This
effect has been termed catalysis \cite{Jo99}, since the state
$|\eta\rangle_{ab}$ is necessary for the process to occur, even
though it is not consumed.

In this Letter we present a novel catalysis effect through
quantum entanglement. A maximally entangled state will be used,
but not consumed, to perform a non-local task that cannot be
achieved without it. The task consists of implementing a certain
two--qubit unitary gate when only some other one is available.
Remarkably, this catalysis is achieved using only local unitary
manipulations. The same construction allows to simulate with a
given non--local interaction other kinds of interactions, which
otherwise could not be simulated using only LOCC. In our method
unitarity of the local manipulations is an important feature,
since it makes possible that some LOCC-inequivalent interactions
become fully equivalent in presence of entanglement. This
sharply contrasts with the case of entangled state conversions
through LOCC manipulations \cite{Jo99}, where LOCC-inequivalent
states must remain inequivalent through catalysis, because the
local measurements needed in the conversions unavoidably
decrease the entanglement between the parties. Another
consequence of our results is that certain Hamiltonians become
equivalent under asymptotic LOCC, a phenomenon that shares
analogies with the one that occurs in transformations between
pure states \cite{Be96}.

Let us consider two parties, Alice and Bob, each of them
possessing a qubit, $A$ and $B$, respectively. Their goal is to
apply certain unitary operator $\tilde U$ to the qubits.
However, they only have at hand another particular two--qubit
unitary operator $U$, and the ability to perform one of the
following classes of operations. (a) LU: local unitary
operations on each qubit; (b) LU+anc: each of the local unitary
operations is jointly performed on a local ancilla, initially in
a product state, and a qubit; (c) LO: each party can perform
general local operations on its qubit (and ancilla); (d) LOCC:
the same as LO but classical communication is also allowed; (e)
cat--LU: the same as LU+anc, but now Alice's and Bob's ancillas
are initially in an entangled state, which can be used, but not
consumed, during the process. Clearly, everything that can be
done in the LU, LU+anc, and LO scenarios, can be also done in
the LOCC scenario. Here we will show that there are operators
$\tilde U$ that cannot be applied in the LOCC scenario, but that
can be achieved in the cat--LU one.

Let $U$ denote a unitary operator acting on two qubits $A$ and
$B$. Using the results of Ref.\ \cite{Kr01}, we can always write
$U^{AB}=(u^A v^B) [U^{AB}_s(c_1,c_2,c_3)] (\tilde u^A
\tilde v^B)$, where
\bma
\bea
\label{U0}
U^{AB}_s(c_1,c_2,c_3)&=& e^{-i \sum_{k=1}^3 c_k
\sigma_k^A\sigma_k^B},\\
\label{restriction}
&& \pi/4\ge c_1 \ge c_2 \ge |c_3|,
\eea
\ema
the $\sigma$'s are Pauli operators, and the $u$'s and $v$'s are
local unitary operators. The superscripts accompanying each
operator indicate the system(s) on which it acts. The
coefficients $c$ can be easily determined using the method
described in Ref.\ \cite{Kr01}. Any two unitary operators are
equivalent under LU (i.e. they can perform the same tasks if
arbitrary local unitary operations on $A$ and $B$ are allowed
before and after their action) if and only if they give rise to
the same $U_s(c_1,c_2,c_3)$. Since in all what follows we will
always allow for LU, we can restrict ourselves to unitary
operators $U$ of the form (\ref{U0}).

In the catalytic scenario, cat--LU, we have at our disposal two
ancillas (qubits) $a$ and $b$, initially in the Bell state
$|B_{0,0}\rangle_{ab}$ \cite{notBell}. We must impose that after
the whole process the ancillas $a$ and $b$ end up again in state
$|B_{0,0}\rangle_{ab}$. We allow for joint unitaries acting on
$A$ and $a$, as well as joint unitaries acting on $B$ and $b$.
We will show that in this situation we can use
$U_s(c_1,c_2,c_3)$ to implement $U_s(c_1+c_2,0,0)$. Later on we
will show that this cannot be achieved without the entangled
ancillas, even if LOCC are allowed.

The above claim about what can be done with $U_s$ in the cat--LU
scenario follows directly from the fact that
\bea
& &\left(w^{Aa}w^{Bb}\right)^\dagger
\left[U^{AB}_s(c_1,c_2,c_3)\right]
\left(w^{Aa} w^{Bb}\right)|\Psi\rangle_{AB}|B_{0,0}\rangle_{ab}
\nonumber\\
\label{catal}
&&=e^{ic_3} \left[U^{AB}_s(c_1+c_2,0,0)\right]
|\Psi\rangle_{AB}|B_{0,0}\rangle_{ab},
\eea
for all $|\Psi\rangle$. Here, the unitary operators $w$ are
defined according to $w|i,j\rangle = |j,i\oplus j\rangle$, and
therefore correspond to a swap operation followed by a c--NOT.
Even though Eq.\ (\ref{catal}) can be directly checked, we will
indicate here the main idea behind this equation. The operators
in the form $U_s$ are diagonal in the Bell basis \cite{notBell},
i.e.
\bma
\bea
U_s(c_1,c_2,c_3) |B_{0,0}\rangle &=& e^{-i(c_1+c_2-c_3)}
|B_{0,0}\rangle,\\ U_s(c_1,c_2,c_3) |B_{1,0}\rangle &=&
e^{-i(c_1-c_2+c_3)} |B_{1,0}\rangle,\\ U_s(c_1,c_2,c_3)
|B_{0,1}\rangle &=& e^{i(c_1+c_2+c_3)} |B_{0,1}\rangle.\\
U_s(c_1,c_2,c_3) |B_{1,1}\rangle &=& e^{-i(-c_1+c_2+c_3)}
|B_{1,1}\rangle.
\eea
\ema
In particular,
\bma
\bea
e^{ic_3}U_s(c_1+c_2,0,0)|B_{\alpha,0}\rangle &=& e^{- i
(c_1+c_2-c_3)}|B_{\alpha,0}\rangle,\\
e^{ic_3}U_s(c_1+c_2,0,0)|B_{\alpha,1}\rangle &=& e^{ i
(c_1+c_2+c_3)}|B_{\alpha,1}\rangle,
\eea
\ema
for $\alpha=\pm 1$. Thus, we see that if we could transform
$|B_{\alpha,\beta}\rangle_{AB} \to |B_{0,\beta}\rangle_{AB}$
before acting with $U_s(c_1,c_2,c_3)$ and then we would invert
such transformation, we would obtain the desired result.
Unfortunately, there exist no such a transformation since two
states ($\alpha=0,1$) have to be mapped onto a single one, and
then back. However, this can be accomplished with the help of
the entangled ancillas, and this is precisely what the operator
$w_{Aa} w_{Bb}$ does: it transforms
$|B_{\alpha,\beta}\rangle_{AB}|B_{0,0}\rangle_{ab} \to
|B_{0,\beta}\rangle_{AB}
|B_{\overline\alpha,\beta}\rangle_{ab}$.

Now, let us show that $U_s(c_1+c_2,0,0)$ cannot be obtained with
the help of $U_s(c_1,c_2,c_3)$ and LOCC for a range of values of
the parameters $c$. Note that this automatically implies that
this task is not possible either with LU, LU+anc, or LO. In the
LOCC scenario we may use two ancillas $a$ and $b$, with
corresponding Hilbert spaces of arbitrary dimensions. The LOCC
consist of generalized measurement on $A$ and $a$, and on $B$
and $b$ involving classical communication before and also after
the application of $U_s(c_1,c_2,c_3)$.

We want that the whole procedure involving a set of LOCC,
followed by the action of $U_s(c_1,c_2,c_3)$, and again another
set of LOCC, reproduce the action of $U_s(c_1+c_2,0,0)$ on any
input state of $A$ and $B$. In particular, we can take $A$ and
$B$ initially entangled with two other, remote qubits $C$ and
$D$, in state
\be
|\Psi_0\rangle_{ABCDab}\equiv |B_{0,0}\rangle_{AC}
|B_{0,0}\rangle_{BD}
 |0\rangle_a |0\rangle_b.
\ee
Let us assume that a set of LOCC takes place {\em before}
$U_s(c_1,c_2,c_3)$ acts. We will now show that one can
substitute these LOCC by local unitaries acting on $A$ and $a$,
and $B$ and $b$. We will use the fact that the whole process
must be described by a unitary operator [$U_s(c_1+c_2,0,0)$]
acting on $A$ and $B$, which implies that the entanglement
between the qubit $C$ ($D$) and the rest of the systems must be
preserved, i.e. the final state must be a maximally entangled
state between $C$ ($D$) and the rest. For a set of outcomes
$\Gamma$ of the generalized measurements performed on $A$ and
$a$, and on $B$ and $b$, before the application of
$U_s(c_1,c_2,c_3)$ we will have that the state of the systems
will change according to $x^{Aa}_{\Gamma} y^{Bb}_{\Gamma}
|\Psi_0\rangle_{ABCDab}$, where $x_{\Gamma}$ and $y_{\Gamma}$
are two operators that depend on the set of outcomes of the
measurements. Let us consider first the action of $x$ (we will
omit the subscript $\Gamma$ in order to keep the notation
readable)
\be
x|0\rangle_A|0\rangle_a = d_{0}|\psi_{0}\rangle_{Aa},\quad
x|1\rangle_A|0\rangle_a = d_{1}|\psi_{1}\rangle_{Aa},
\ee
where $|\psi_{0,1}\rangle$ are normalized states. Note that it
can occur neither that $|d_{0}|\ne|d_{1}|$ nor that
$|\psi_{0}\rangle$ and $|\psi_{1}\rangle$ are not orthonormal.
If this were the case, then the entanglement of the qubit $C$
with the rest of the systems would decrease. According to well
known results on entanglement concentration \cite{Lo01}, this
entanglement cannot be recovered later on with the help of LOCC.
Since the whole protocol does not involve joint actions with
remote qubit C, this immediately would contradict the fact that
this entanglement has to be maintained at the very end of the
process. Thus, we must have that $|d_{0}|=|d_{1}|\equiv d$ and,
at the same time, $|\psi_{0}\rangle$ and $|\psi_{1}\rangle$ are
orthonormal. But in this case we can always find certain unitary
operator $u$ acting on $A$ and $a$ such that $du$ gives the same
action as $x$ on the relevant states. Thus, we can substitute
$x_\Gamma$ by a unitary operator $u_\Gamma$ chosen randomly with
probability $|d_{\Gamma}|^2$. The same analysis applies to
$y_\Gamma$.

According to this result, the problem reduces to showing that
\be
\label{Phi1}
|\Phi_1(\Psi)\rangle\equiv\left[U^{AB}_s(c_1+c_2,0,0) \right]
|\Psi\rangle_{AB}|0,0\rangle_{ab},
\ee
cannot be obtained starting from
\be
|\Phi_2(\Psi)\rangle\equiv\left[U^{AB}_s(c_1,c_2,c_3)\right]
\left(x^{Aa} y^{Bb}\right) |\Psi\rangle_{AB}|0,0\rangle_{ab},
\ee
using LOCC, for all $|\Psi\rangle$ and where $x$ and $y$ are
unitary. In order to prove that, we restrict the values of the
parameter $c$ to satisfy $c_3=0$, $c_2>0$, and $c_1+c_2\le
\pi/4$, and use the following fact \cite{Ni99}: if $|\Psi_1\rangle$
can be obtained by LOCC out of $|\Psi_2\rangle$, then
\be
\label{Nielsen}
P(\Psi_1)\ge P(\Psi_2),
\ee
where
\be
P(\Psi)\equiv\max_{||\psi||=||\phi||=1} |\langle\psi|\langle
\phi||\Psi\rangle|^2.
\ee
[$P$ is the square of the maximal Schmidt coefficient.] In
particular, if we take in (\ref{Phi1})
$|\Psi_{i,j}\rangle_{AB}=|i\rangle_A|j\rangle_B$ ($i,j=0,1$), we
have that $P[\Phi_1(\Psi_{i,j})]=\cos^2(c_1+c_2)$. Defining
\bma
\bea
|\psi_i\rangle_{Aa} &\equiv& x^{Aa} |i,0\rangle_{Aa},\\
|\varphi_j\rangle_{Bb} &\equiv& y^{Bb} |j,0\rangle_{Bb},
\eea
\ema
we will show that it is not possible to have
\be
\label{cond}
\left| \langle \psi_i|\langle \varphi_j| |\Phi_2(\Psi_{i,j})\rangle
\right|^2
\le \cos^2(c_1+c_2),
\ee
for all $i,j=0,1$, and therefore that condition (\ref{Nielsen})
is violated. We can always write
\be
\label{psivarphi}
|\psi\rangle_{Aa}|\varphi\rangle_{Bb} =
\sum_{\alpha,\beta=0,1} |B_{\alpha,\beta}\rangle_{AB}
|N_{\alpha,\beta}\rangle_{ab},
\ee
where the $n_{\alpha,\beta}\equiv||N_{\alpha,\beta}||^2\ge 0$
add up to one. Thus, condition (\ref{cond}) reduces to
\bea
&&\left| e^{-i(c_1+c_2)} n_{0,0} + e^{i(c_1+c_2)} n_{0,1}
\right.\nonumber\\
&& \left.+ e^{-i(c_1-c_2)} n_{1,0} + e^{i(c_1-c_2)}
n_{1,1}\right|^2
\le
\cos^2(c_1+c_2).
\eea
Actually, it can be easily shown that the left hand side is
always larger or equal than the right hand side, the equality
holding only for $n_{1,0}=n_{1,1}=0$ and $n_{0,0}=n_{0,1}=1/2$.
Using these results in Eq.\ (\ref{psivarphi}) and imposing that
$|\psi_i\rangle_{Aa}|\varphi_j\rangle_{Bb}$ is a product state,
we obtain that it must be of either of the form
$|0,1\rangle_{AB}|\mu_i,\nu_j\rangle_{ab}$ or
$|1,0\rangle_{AB}|\mu_i,\nu_j\rangle_{ab}$. Now, recalling that
$|\psi_i\rangle_{Aa}|\varphi_j\rangle_{Bb}$ must be created
using local unitary operators acting on $A$ and $a$, and $B$ and
$b$ out of $|i,0\rangle_{Aa}|j,0\rangle_{Bb}$ one readily finds
that this is impossible for all $i,j=0,1$. Thus, we have proven
that $U_s(c_1+c_2,0,0)$ cannot be obtained with the help of
$U_s(c_1,c_2,0)$ and LOCC for $\pi/4\ge c_1+c_2>0$ and $c_1\ge
c_2>0$.

In the following, we will analyze the implications of our
catalytic method in the context of infinitesimal transformations
of two-qubits \cite{Du00,Nielsen,Beth,Be01,Vi01}. Remarkably,
the study of this kind of transformations has allowed to
establish a partial order in the set of all possible physical
interactions (or Hamiltonians) \cite{Be01}. This partial order
is related to whether a given interaction can {\em simulate}
(i.e., produce the same results of) another one, when certain
operations are allowed. In this context, the necessary and
sufficient conditions for a two-qubit Hamiltonian $H$ to be able
to simulate another $H'$ under LU, LU+anc and LOCC have been
derived \cite{Be01,Vi01}, giving the same conditions. One can
immediately see from our general results on unitary operators
that in the catalytic scenario, these conditions are relaxed,
i.e. there are certain Hamiltonians that can simulate other
under cat--LU, but not under LOCC. Here we will analyze this
fact in detail and extract some conclusions.

Thus, we consider $U=e^{-iH\delta t}$, where $H=H^\dagger$ is a
Hamiltonian acting on the qubits $A$ and $B$ and $||H\delta t||
\ll 1$. Again, since we allow for arbitrary local unitaries
at any time, we can restrict ourselves to Hamiltonians of the
form
\bma
\bea
H(c_1,c_2,c_3) &=& \sum_{k=1}^3 c_k
\sigma_k^A \sigma_k^B,\\ c_1\ge c_2\ge |c_3|.
\eea
\ema
In Refs.\ \cite{Be01,Vi01} it has been shown that given
$H(c_1,c_2,c_3)$, a total time $\delta t$, and if we allow for
LOCC after time steps smaller than $\delta t$, then we can
obtain the operation generated by $H(\tilde c_1,\tilde
c_2,\tilde c_3)$ during the same time $\delta t$ up to second
order corrections in $H\delta t$ if and only if
\bma
\label{conditionsinf}
\bea
c_1 + c_2 -c_3 &\ge& \tilde c_1 + \tilde c_2 -\tilde c_3,\\
\label{cond2}
 c_1 &\ge& \tilde c_1,\\
 c_1 + c_2 + c_3 &\ge& c_1 + c_2 + c_3.
\eea
\ema
This implies that under LOCC, $H$ can simulate $\tilde H$ if and
only if these conditions are satisfied.

If we use our catalytic method, we have that it is possible to
simulate $\tilde H(c_1+c_2,0,0)$ with $H(c_1,c_2,c_3)$, which
for $c_2\ne 0$ violates condition (\ref{cond2}). In fact, taking
$c_3=0$, we see that $H_1 \equiv H(c_1+c_2,0,0)$ can simulate
$H_2 \equiv H(c_1, c_2,0)$ as well, since conditions
(\ref{conditionsinf}) are fulfilled. Thus, our catalytic method
makes any pair of Hamiltonians of the form $H_1$ and $H_2$
equivalent, although they are inequivalent under LOCC
simulation. This result also has fundamental implications in the
study of {\em asymptotic} simulation of interactions using
LU+anc. There $N$ applications of an evolution generated by $H$
for a time $\delta t$ are available, in the limit $\delta t\to
0$ and $N\delta t\to
\infty$. $H_1$ can simulate $H_2$ even for finite $N$ \cite{Be01,Vi01}.
We can now use $H_2$ for $N_0$ times to create a maximally
entangled state of the ancillas \cite{Du00} with $N_0 \delta t$
finite, which could then be used to catalyze the Hamiltonian
evolution generated by $H_1$ a number $N-N_0\sim N$ of times.

So far, we have seen that under the catalytic scenario, some
Hamiltonians acting on two qubits become equivalent. Of course,
an important question is whether all Hamiltonians become
equivalent in that scenario \cite{Bennetpc}. We now show that
this is not the case. We derive a set of necessary conditions
similar to (\ref{conditionsinf}) that the Hamiltonians $H$ and
$\tilde H$ must fulfill for $H$ to be able to simulate $\tilde
H$. First, we will use that both Hamiltonians are diagonal in
the Bell basis \cite{notBell}, and we will call the
corresponding eigenvalues
\bma
\bea
\lambda_1= c_1+c_2-c_3, &&\quad \tilde \lambda_1= \tilde c_1+\tilde
c_2-\tilde c_3 + \tilde c_4,\\
\lambda_2= c_1-c_2+c_3, &&\quad \tilde \lambda_2= \tilde c_1-\tilde
c_2+\tilde c_3 + \tilde c_4,\\
\lambda_3= -c_1+c_2+c_3, &&\quad \tilde \lambda_3= -\tilde c_1+\tilde
c_2+\tilde c_3 + \tilde c_4,\\
\lambda_4= -c_1-c_2-c_3, &&\quad \tilde \lambda_4= -\tilde c_1-\tilde
c_2-\tilde c_3 + \tilde c_4.
\eea
\ema
Note that with these numeration, the $\lambda$'s and $\tilde
\lambda$'s are sorted in decreasing order. We have also taken
into account a global constant $\tilde c_4$, since it will be
important in the discussion below. We will show that if $H$ can
simulate $\tilde H$ under cat--LU, then
\bma
\bea
\label{cond21}
c_1+ c_2-c_3 &\ge& \tilde c_1+\tilde c_2-\tilde c_3 + \tilde
c_4,\\
\label{cond22}
c_1+ c_2+c_3 &\ge& \tilde c_1+\tilde c_2+\tilde c_3 -\tilde
c_4,\\
\label{cond23}
\sum_{k=1}^3 |c_k| &\ge& \sum_{k=1}^4 |\tilde c_k|.
\eea
\ema
These conditions mean, for example, that with $H(c_1,c_2,c_3)$
it is not possible to efficiently simulate either $\tilde
H(c_1+c_2+c_3,0,0)$ ---which would imply catalytic equivalence
of all interactions since the converse simulation is possible
[cf. (\ref{conditionsinf})]---, nor $H(c_1,c_2,-c_3)$
---which excludes the simulation of a time-reversed evolution of
$H(c_1,c_2,c_3)$.

Following the same steps as in \cite{Vi01} we find that $H$ can
efficiently simulate $\tilde H$ using LOCC only if there exists
a set of unitary operators $u_m$ and $v_m$ and some positive
numbers $p_m$ which add up to one, such that
\bea
\label{inter}
&&\sum_{m}
p_m(u_m^{Aa}v_m^{Bb})^\dagger\left[H^{AB}\otimes\one^{ab}\right]
(u_m^{Aa}v_m^{Bb})
|\Psi\rangle_{AB}|\Phi_0\rangle_{ab}\nonumber\\
 && =\left[\tilde
H^{AB}\otimes\one^{ab}\right]|\Psi\rangle_{AB}|\Phi_0\rangle_{ab},
\eea
for all $|\Psi\rangle$ and certain fixed state $|\Phi_0\rangle$
of arbitrary dimensional ancillas. Here we have included
$\one^{ab}$ to make the formula more explicit. According to a
basic result in the theory of majorization \cite{Uhxx}, the
operator resulting from the sum over $m$ in (\ref{inter}) must
have the eigenvalues lying in the interval
$[\lambda_1,\lambda_4]$. This automatically implies that the
operator $\tilde H^{AB}$ must also have its eigenvalues in the
same interval, which leads to $\lambda_1\ge \tilde \lambda_1$
and $\lambda_4\le\tilde \lambda_4$, and therefore to
(\ref{cond21},\ref{cond22}). In order to obtain the last
condition (\ref{cond23}), we apply the bra $_{ab}\langle
\Phi_0|$ to both sides of Eq.\ (\ref{inter}), multiply the
corresponding equation by $\sigma_k^A\sigma_k^B/4$ and trace
with respect to $A$ and $B$. Taking the absolute values of the
resulting expressions, and adding from $k=1,\ldots,4$ we obtain
\be
\label{ineq}
\sum_{k=1}^4 |\tilde c_k| \le \sum_{n=1}^3 |c_n| \sum_m p_m h_{n,m},
\ee
where
\be
h_{n,m} = \frac{1}{4}\sum_{k=1}^4
|_{ab}\langle\Phi_0|X_{k,n,m}^a Y_{k,n,m}^b|\Phi_0\rangle_{ab}|,
\ee
and
\bma
\bea
X_{k,n,m}^a &=& {\rm tr}_A[\sigma_k^A (u_m^{Aa})^\dagger
\sigma_m^A u_m^{Aa}].\\
 Y_{k,n,m}^b &=& {\rm tr}_B[\sigma_k^B (v_m^{Bb})^\dagger \sigma_m^B
 v_m^{Bb}].
\eea
\ema
Using Cauchy--Schwarz inequality, we have
\bea
h_{n,m} &\le& \left[\frac{1}{4}\sum_{k=1}^4
 \langle\Phi_0|X^{a}_{k,n,m}
 (X^{a}_{k,n,m})^\dagger|\Phi_0\rangle\right]^{1/2}\nonumber\\
 &\times& \left[\frac{1}{4}\sum_{k=1}^4
 \langle\Phi_0|Y^{b}_{k,n,m}
 (Y^{b}_{k,n,m})^\dagger|\Phi_0\rangle\right]^{1/2}=1,\nonumber
 \eea
where for the last equality we have used the fact that
$\sigma_k$ form an orthonormal basis in the space of operators
acting on a qubit. Substituting $h_{n,m}\le 1$ in Eq.\
(\ref{ineq}), we finally obtain condition (\ref{cond23}).

In conclusion, we have shown that certain unitary operations can
be catalyzed by an entangled state, in the sense that the state
is not consumed but without it the process would not be
possible. We have also shown that the method introduced here
allows to make equivalent certain kind of interactions acting on
two qubits. This fact allows for these interactions to become
equivalent in the asymptotic limit, which is compatible with the
conjecture that all two-qubit interactions are equivalent in the
asymptotic limit \cite{Bennetpc}.

We thank C. H. Bennett for stimulating discussions. This work
was supported by the Austrian Science Foundation under the SFB
``control and measurement of coherent quantum systems'' (Project
11), the European Community under the TMR network
ERB--FMRX--CT96--0087, project EQUIP (contract IST-1999-11053),
and contract HPMF-CT-1999-00200, the European Science
Foundation, and the Institute for Quantum Information GmbH.

% -------------------------------------------------------------


\begin{references}
\bibitem{notBell}
We denote the Bell states as follows:
\bma
\bea
 |B_{0,0}\rangle &\equiv& \frac{1}{\sqrt{2}} (|0,1\rangle
 +|1,0\rangle),~
 |B_{1,0}\rangle \equiv \frac{1}{\sqrt{2}}(|0,0\rangle + |1,1\rangle),
\nonumber\\
 |B_{0,1}\rangle &\equiv&\frac{1}{\sqrt{2}} (|0,1\rangle -
 |1,0\rangle),~
 |B_{1,1}\rangle \equiv\frac{1}{\sqrt{2}} (|0,0\rangle -
|1,1\rangle).\nonumber
\eea
\ema
\end{references}
\end{document}